\documentclass[showpacs,preprintnumbers,amsmath,amssymb]{revtex4}

\usepackage{epsfig}
\usepackage{graphicx}
\usepackage{dcolumn}
\usepackage{bm}

\begin{document}

\title{

Multiscale analysis of heart rate, blood pressure and respiration time series}
\author{L.Angelini$^{1,2,3}$, R. Maestri$^4$, D. Marinazzo$^{1,2,3}$, L. Nitti$^{1,3,5}$, M. Pellicoro$^{1,2,3}$,
G. D. Pinna$^4$,  S. Stramaglia$^{1,2,3}$, S.A. Tupputi$^2$}
 \affiliation{$^1$TIRES-Center
of
Innovative Technologies for Signal Detection and Processing,\\
Universit\`a di Bari, Italy\\
$^2$ Dipartimento Interateneo di Fisica, Italy \\
$^3$Istituto Nazionale di Fisica Nucleare, Sezione di Bari, Italy \\
$^4$Dipartimento di Bioingegneria, Fondazione S. Maugeri, IRCCS, Istituto Scientifico di
Montescano (PV), Italy \\
$^5$Dipartimento di Biochimica Medica, Biologia Medica e Fisica Medica, University of
Bari, Italy}

\date{\today}

\begin{abstract}
We present the multiscale entropy analysis of short term physiological time series of
simultaneously acquired samples of heart rate, blood pressure and lung volume, from
healthy subjects and from subjects with Chronic Heart Failure. Evaluating the complexity
of signals at the multiple time scales inherent in physiologic dynamics, we find that
healthy subjects show more complex time series at large time scales; on the other hand,
at fast time scales, which are more influenced by respiration, the pathologic dynamics of
blood pressure is the most random. These results robustly separate healthy and pathologic
groups. We also propose a multiscale approach to evaluate interactions between time
series, by performing a multivariate autoregressive modelling of the coarse grained time
series: this analysis provides several new quantitative indicators which are
statistically correlated with the pathology.

\end{abstract}

\maketitle

\section{Introduction}
Physiological systems are ruled by mechanisms operating across multiple temporal
scales. Many approaches have been developed in the last years to analyze these complex
signals, including, for example, studies of: Fourier spectra \cite{akselrod}, chaotic
dynamics \cite{bablo}, scaling properties \cite{nunes}, multifractal properties
\cite{ivanov}, correlation integrals \cite{lehnertz}, $1/f$ spectra \cite{peng} and
synchronization properties \cite{tass}. A recently proposed approach, multiscale entropy
analysis (MSE) \cite{costa}, compares the degree of complexity of time series at varying
temporal scale, and has been applied to 24 hours electrocardiographic recordings of
healthy subjects, subjects with congestive heart failure, and subjects with atrial
fibrillation. Results from this analysis support the general ${\it complexity-loss}$
theory of aging and disease, since healthy and young dynamics are the most complex.

In this paper we apply the MSE analysis to short-term simultaneous recordings of
electrocardiogram, respiration signal and arterial blood pressure, from healthy subjects
and from subjects with Chronic Heart Failure (CHF), a disease associated with major
abnormalities of autonomic cardiovascular control.

We also consider here a multiscale version of the classical multivariate autoregressive
analysis of time series, to find scale-dependent patterns of interactions between the
physiological time series here considered. The paper is organized as follows. In the next
section we describe our data set, the methods and the results we obtain. Some conclusions
are drawn in section III.

\section{Data, methods and results}
We briefly recall the MSE method \cite{costa}. Given a one-dimensional discrete time
series, consecutive coarse grained time series, corresponding to scale factor $\tau$, are
constructed in the following way. First, the original time series is divided into
nonoverlapping windows of length $\tau$; then, data points inside each window are
averaged, so as to remove fluctuations with time scales smaller than $\tau$. For scale
one, the coarse grained time series is simply the original time series; the length of
each coarse grained time series is equal to the length of the original time series
divided by the scale factor $\tau$. Finally an entropy measure $S_E$ is calculated for
each coarse grained time series and plotted as function of the scale factor $\tau$. $S_E$
coincides with the parameter $S_E (m,r)$, introduced by Richman and Moorman
\cite{complex} and termed {\it sample entropy}, which is related to the probability that
sequences from the time series, which are close (within $r$) for $m$ points, remain close
at the subsequent data point. In the original proposal both the sequence length $m$ and
the tolerance parameter $r$ were kept fixed as $\tau$ was varied, so that changes in
$S_E$ on each scale were depending both on the regularity and the variability of the
coarse grained sequences \cite{comment}. In the present work we take $r$, at each $\tau$,
inversely proportional to the standard deviation (SD) of the coarse grained time series,
and consider separately how the SD of signals varies with the time scale.

Our data are from 47 healty subjects and 275 stable mild to moderate CHF patients in
sinus rhythm admitted to the Heart Failure Unit of the Scientific Institute of Montescano
for evaluation and treatment of heart failure, usually in conjunction with evaluation for
heart transplantation. Concerning the second group, cardiac death occurred in 54 patients
during a 3-year-follow-up. In two different conditions of respiration, basal and paced
breathing (at 0.25 Hz) \cite{paced}, ten minutes long physiological recordings have been
made on these subjects, leading to four time series. Firstly, the heart RR interval time
series ({\it rri}); for each cardiac cycle, corresponding values of the systolic arterial
pressure {\it sap}, the diastolic arterial pressure {\it dap} and the instantaneous lung
volume {\it ilv} were computed. The four time series have then been re-sampled at 2Hz
using a cubic spline interpolation. Part of this data set (the {\it sap} time series) has
been already analyzed in \cite{ancona} using a different approach.

In figure 1 we depict the standard deviations of the coarse grained time series in basal
condition. Due to the short length of the samples at our disposal, we consider $\tau \le
10$ so as to have sufficient statistics at each scale; this implies that our analysis
will be limited to part of the High Frequency (HF) band (0.15-0.45Hz), the band in which
the respiratory rhythm of most people lies. In all cases, on average the standard
deviation is a decreasing function of the scale; healthy subjects show greater
variability than patients, except for {\it ilv} signals, where patients on average have
the highest variability. Similar patterns of standard deviations are obtained in paced
breathing conditions.

As already stated, to extract the sample entropy from these signals, we take $r$ equal to
a fixed percentage ($15\%$) of the standard deviations of the coarse grained time series;
we take $m=1$. In figure 2 we depict the average $S_E$ of {\it rri} time series of
controls, patients and dead patients, in basal condition (high) and paced breathing
(low). Concerning the basal case, we note that controls have always significantly higher
entropy than CHF patients, at all scales, and that dead patients show slightly more
regular {\it rri} time series than the average over all patients. The severity of the
pathology seems to be correlated with the loss of entropy. On the right we depict, as a
function of the scale factor $\tau$, the probability that {\it rri} entropy values from
controls and patients were drawn from the same distribution, evaluated by non parametric
rank sum Wilcoxon test: the discrimination is excellent at intermediate $\tau$'s. This
picture is in agreement with findings in \cite{costa}, corresponding to controls and
subjects with congestive heart failure in sinus rhythm, except for a different form of
the entropy curve for patients, which indeed depends on the pathology. In the case of
paced breathing the three curves get closer and the discrimination, between patients and
controls, reduces: paced breathing seems, in the case of {\it rri} entropy, to reduce
differences between patients and controls.

In figure 3 we depict $S_E$ of {\it sap} time series. We find that at low $\tau$ patients
have higher entropy, whilst at large $\tau$ they have lower entropy than controls. The
crossover occurs at $\tau=3$ in basal conditions, and $\tau\sim 6$ for paced breathing.
The ${\it complexity-loss}$ paradigm, hence,  here holds only for large $\tau$. This may
be explained as an effect of respiration, whose influence seems to become weaker as
$\tau$ increases. This effect is more evident in conditions of paced breathing. Our
results are consistent with those obtained in \cite{ancona} using a different approach
and with $\tau =1$. It is interesting to observe that curves corresponding to dead
patients are always farther, from the controls curve, than the average curve from all
patients; departure from the controls curve seems to be connected with the severity of
the disease.

In figure 4 we consider {\it dap} time series. We find a similar pattern to {\it sap}:
patients have higher entropy at low $\tau$ and lower entropy than controls at large
$\tau$. Again the crossover occurs at $\tau=3$ in basal conditions, and $\tau=6$ for
paced breathing.

Now we turn to consider {\it ilv} time series, as depicted in figure 5. In the basal
case, controls have higher entropy at small scales. On the other hand controls show lower
entropy than patients at $\tau > 7$: patients pathologically display fluctuations of {\it
ilv} at larger scales than healthy subjects. Under paced breathing, controls are
characterized by reduced fluctuations at high $\tau$; at $\tau =4$, when the window size
is half of the respiration period, controls show a local minimum of the entropy. These
phenomena are not observed for patients, where paced breathing is less effective in
regularizing the {\it ilv} time series.

Next we implement a multiscale version of autoregressive modelling of time series (see,
e.g., \cite{kantz}). For each scale factor $\tau$, we denote ${\bf x}=({\it rri},{\it
sap},{\it dap},{\it ilv})$ the four-dimensional vector of the coarse grained time series.
At each scale, all coarse grained time series are normalized to have unit variance. A
multivariate autoregressive model of unity order is then fitted (by standard least
squares minimization) to data:
\begin{equation}
{\bf x}(t)= A \;{\bf x}(t-1);
\end{equation}
A is a $4\times 4$ matrix, depending on $\tau$, whose element $A_{ij}$ measure the causal
influence of $j-th$ time series on the $i-th$ one. Some of these matrix elements are
found to be significantly different in patients and controls, as described in the
following.

Firstly we consider the interactions between heart rate and blood pressure. In
physiological conditions heart rate and arterial pressure are likely to affect each other
as a consequence of the simultaneous feedback baroreflex regulation from {\it sap-dap} to
{\it rri} and feedforward mechanical influence from {\it rri} to {\it sap-dap}
\cite{koepchen}.

In figure 6 the curves representing the causal relationship ${\it rri}\to {\it sap}$ are
represented. Both in basal and paced breathing conditions, this coefficient is always
negative and is stronger for controls. Two  mechanisms determine the feedforward
influence $rri\to sap$. Firstly the Starling law, stating that when the diastolic filling
of the heart is increased or decreased with a given volume, the volume of blood which is
then ejected from the heart increases or decreases by the same amount. More blood in:
more blood out. This mechanism favors an increase of sap-dap as the rri interval
increases, i.e. a positive coefficient $rri\to sap$. The second mechanism is diastolic
decay, described by the Windkessel model of the capacitative property of arteries; as rri
interval increases, this effect tends to lower sap-dap values and gives a  negative
contribution to the coefficient $rri\to sap$. Our finding suggests that the second
mechanism is dominant. The difference between patients and controls is significant at low
and intermediate $\tau$, and especially in basal conditions. The coefficient ${\it
rri}\to {\it dap}$ shows a behavior very similar to those of  ${\it rri}\to {\it sap}$,
i.e. it is always negative and is stronger for controls.

Evaluation of baroreflex regulation $sap$-$dap$ $\to rri$ is an important clinical tool
for diagnosis and prognosis in a variety of cardiac diseases \cite {pin}. Recent studies,
see e.g. \cite{nollo} and references therein, have suggested that spontaneous
fluctuations of arterial pressure and {\it rri} offer a noninvasive method for assessing
baroreflex sensitivity without use of provocative tests employing injection of a
vasoconstrictive drug or manipulation of carotid baroreceptor. In fig. 7 we depict the
interaction ${\it dap}\to {\it rri}$ as extracted by our approach, showing high
discrimination between controls and patients. In basal conditions this coefficient is
positive for controls and negative for patients. Moreover, this coefficient for patients
is much influenced by respiration: in paced breathing conditions it is almost zero for
patients, while being positive for controls. It is worth stressing that the interaction
${\it dap}\to {\it rri}$, evaluated by the present approach, has only little relation
with the baroreflex sensitivity index considered, e.g., in \cite{nollo}; indeed the
procedures for evaluating these quantities differ in several steps. For example in our
approach all time series are centered and normalized, hence the interaction between
arterial pressure and $rri$ is described only qualitatively.

Human respiration interacts with heart rate, originating the well known phenomenon of
respiratory sinus arrhythmia \cite{hirsch}. We find that the interaction $rri \to ilv$ is
significantly (p $< 10^{-4}$) stronger in controls than patients, under paced breathing
and using $\tau =4$. We also find that the interaction $ilv \to rri$ is positive and
significantly (p $< 10^{-5}$) stronger in controls, in basal conditions and at high
frequencies ($\tau\le 4$).

Let us now turn to consider {\it self interactions} of time series. The matrix element
$A_{11}$ describes how much the {\it rri} signal depends on its value at the previous
time. As it is shown in figure 8, in basal conditions $A_{11}$ is significantly lower for
controls. In paced breathing conditions significant difference is found at high $\tau$.
Also the self interaction of {\it dap} time series gives rise to an interesting pattern.
It is stronger for controls, especially at low $\tau$,  leading to high discrimination
between controls and patients at low $\tau$ as figure 9 shows.

The interaction of systolic and diastolic arterial pressure in healthy subjects has been
recently studied in \cite{winfree}. In the present analysis we find significant
differences between patients and controls when the interaction {\it sap}$\to${\it dap} is
considered, see figure 10. For controls, this coefficient is always negative and its
strength increases with $\tau$.

It is known that respiration interacts in an open loop way with arterial pressure, mainly
through a mechanical mechanism \cite{deboer}. Our findings confirm it; indeed we find no
significant ${\it sap}\to {\it ilv}$ interaction, but significant (p $< 10^{-3}$)
differences between patients and controls are found when the interaction {\it
ilv}$\to${\it sap} is considered: controls show reduced interaction w.r.t. patients.

\section{Conclusions}
In the present paper we have presented the multiscale entropy analysis of short term
physiological time series. We have shown that the analysis of \cite{costa} can be
successfully performed also on short $rri$ recordings, still leading to separation
between controls and patients. Moreover we extend the analysis by considering
simultaneously acquired recordings of {\it sap}, {\it dap} and {\it ilv}. We have also
proposed a multiscale approach to evaluate interactions between time series, by
performing a multivariate autoregressive modelling of the coarse grained time series.
This analysis has put in evidence interesting patterns of interactions between time
series, while providing several new quantitative indicators which are statistically
correlated with the CHF pathology, and which can be employed for diagnosis of CHF
patients. Separating dead patients from alive patients is a very important task, since a
good estimation of the probability of surviving of a given patient would be valuable when
a decision has to be made with respect to the therapy to be undertaken. The separating
performances provided by our indicators in this case are not good as those obtained
separating patients and controls. Further work must be done to deal with the separation
between dead patients and alive patients; in particular it will be interesting to repeat
this analysis with longer recordings so as to take into account fluctuations in lower
frequency bands.

\begin{figure}[ht!]
\begin{center}
\epsfig{file=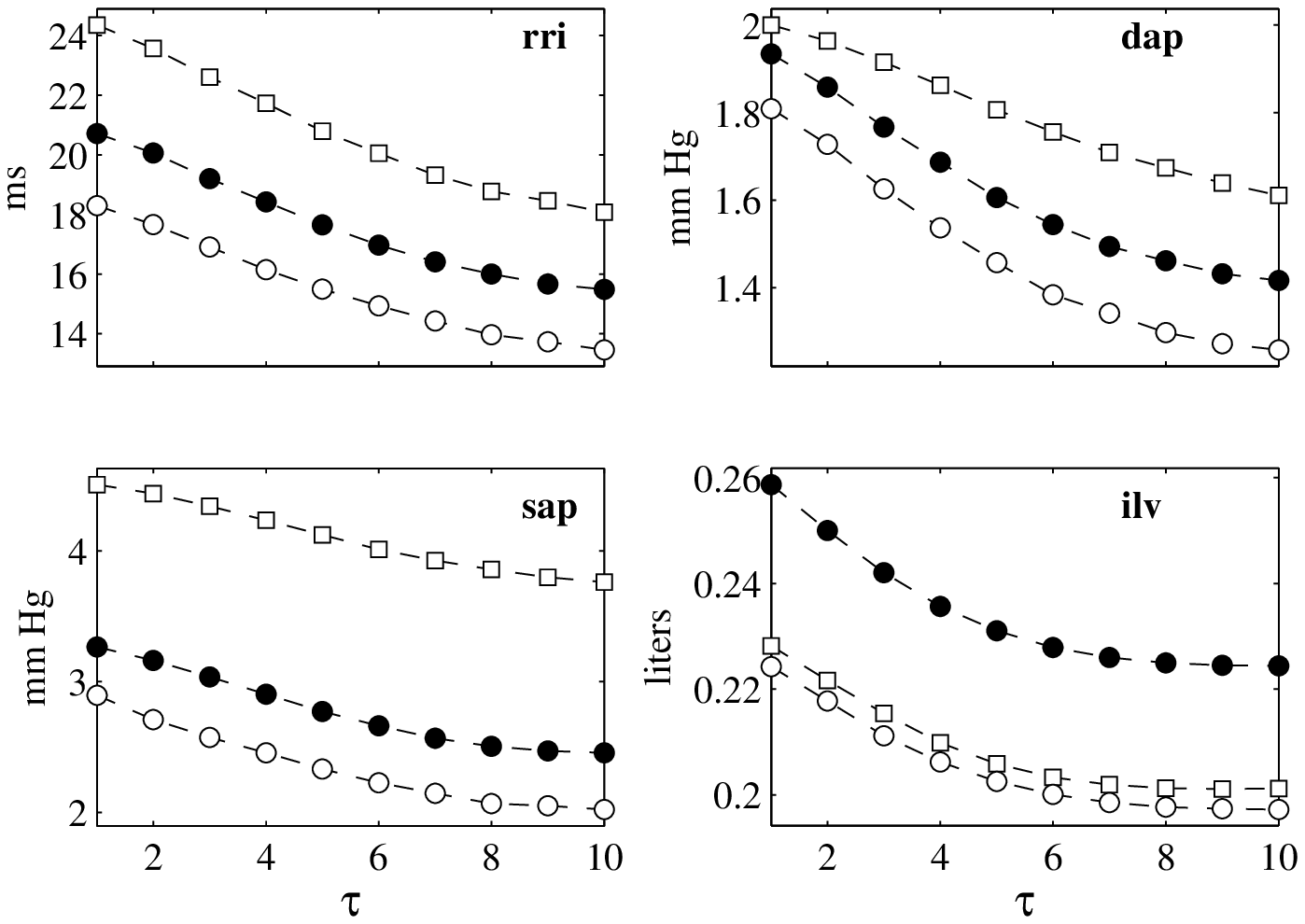,height=8.cm}
\end{center}
\caption{{\small Standard deviations are plotted versus $\tau$ for the coarse grained
time series, in basal condition. Empty squares are the averages over the 47 healthy
subjects, full circles are the averages over the 275 CHF patients, and empty circles are
the averages over the 54 patients for whom cardiac death occurred. Top left: SD of {\it
rri} time series. Top right: SD of {\it dap} time series. Bottom left: SD of {\it sap}
time series. Bottom right: SD of {\it ilv} time series.\label{fig1}}}
\end{figure}

\begin{figure}[ht!]
\begin{center}
\epsfig{file=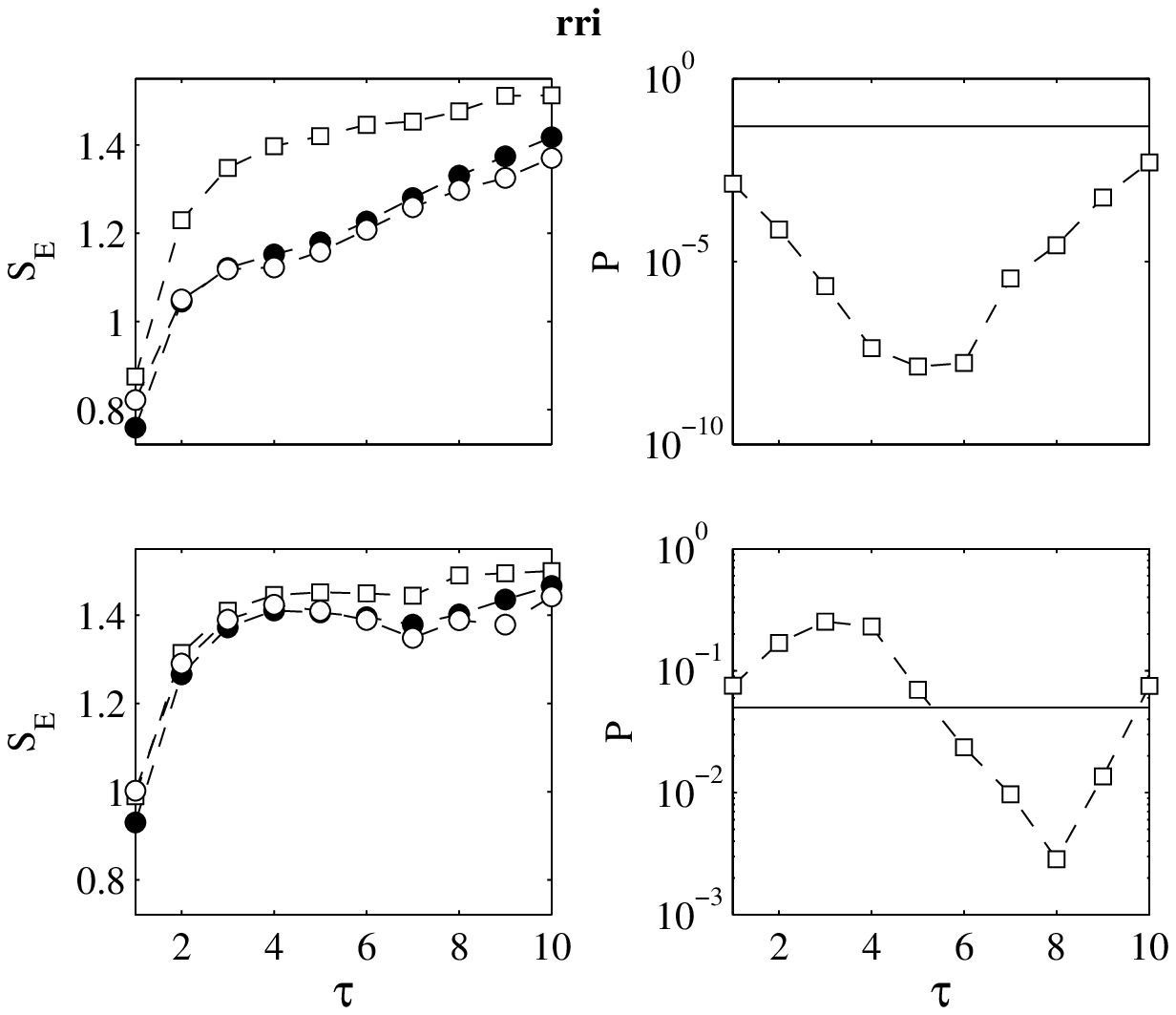,height=8.cm}
\end{center}
\caption{{\small Sample entropy of {\it rri} time series plotted versus $\tau$. Empty
squares are the averages over the 47 healthy subjects, full circles are the averages over
the 275 CHF patients, and empty circles are the averages over the 54 patients for whom
cardiac death occurred. Top left: $S_E$ in basal condition. Top right: the probability
that basal $S_E$ values from controls and patients were drawn from the same distribution,
evaluated by non parametric test. Bottom left: $S_E$ in paced breathing condition. Bottom
right: the probability that paced breathing $S_E$ values from controls and patients were
drawn from the same distribution, evaluated by non parametric test.\label{fig2}}}
\end{figure}

\begin{figure}[ht!]
\begin{center}
\epsfig{file=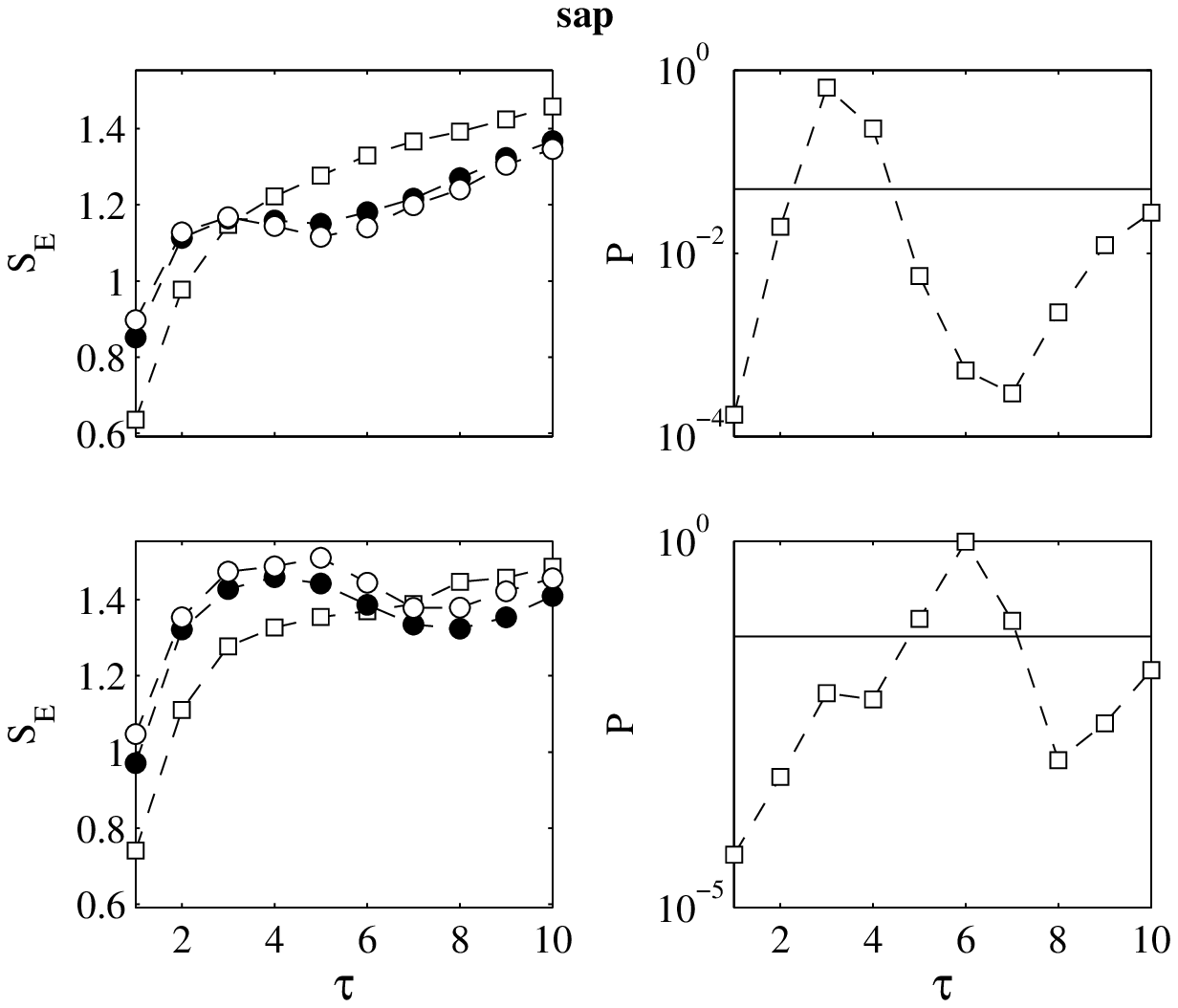,height=8.cm}
\end{center}
\caption{{\small Sample entropy of {\it sap} time series plotted versus $\tau$. Empty
squares are the averages over the 47 healthy subjects, full circles are the averages over
the 275 CHF patients, and empty circles are the averages over the 54 patients for whom
cardiac death occurred. Top left: $S_E$ in basal condition. Top right: the probability
that basal $S_E$ values from controls and patients were drawn from the same distribution,
evaluated by non parametric test. Bottom left: $S_E$ in paced breathing condition. Bottom
right: the probability that paced breathing $S_E$ values from controls and patients were
drawn from the same distribution, evaluated by non parametric test.\label{fig3}}}
\end{figure}

\begin{figure}[ht!]
\begin{center}
\epsfig{file=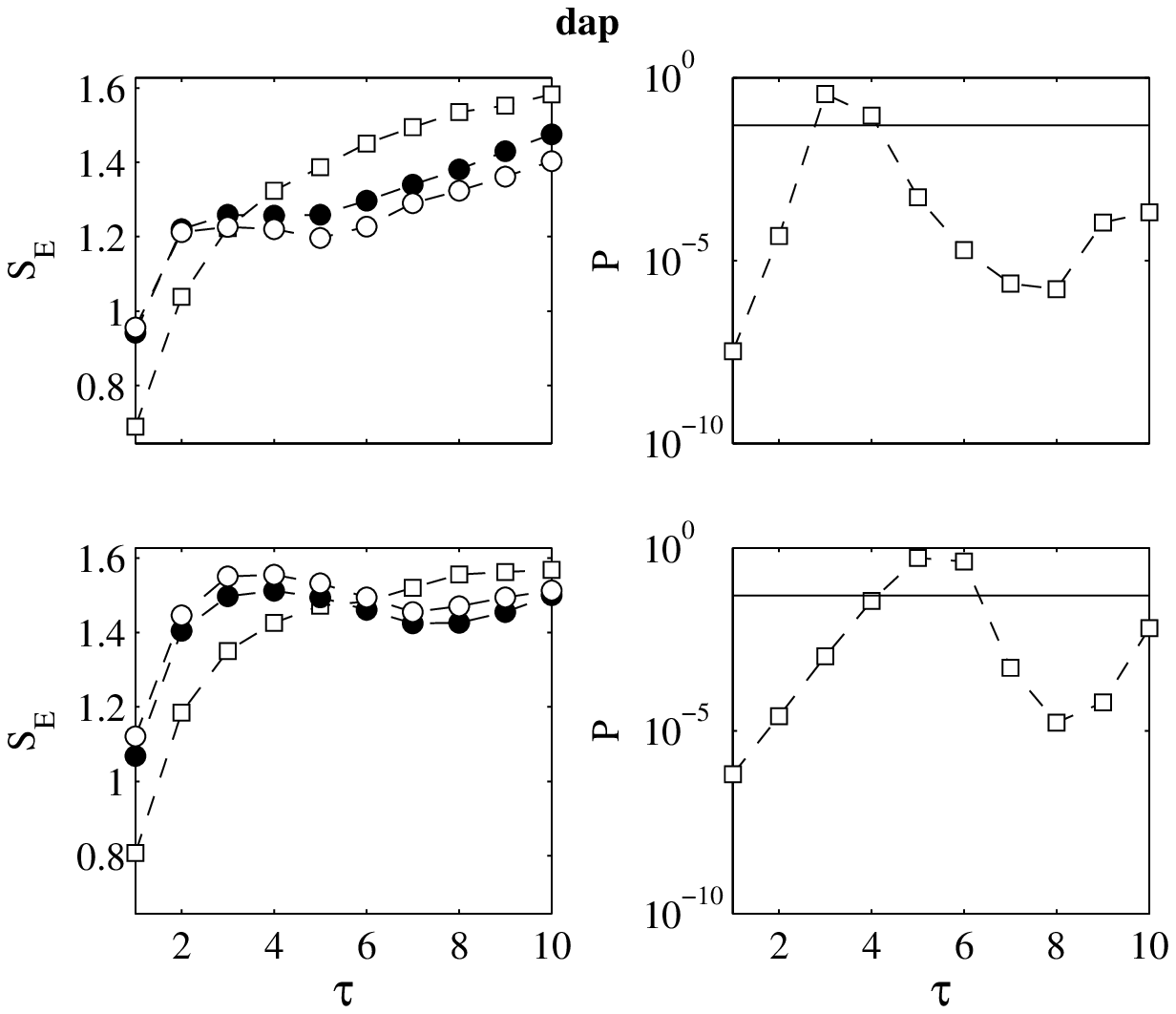,height=8.cm}
\end{center}
\caption{{\small Sample entropy of {\it dap} time series plotted versus $\tau$. Empty
squares are the averages over the 47 healthy subjects, full circles are the averages over
the 275 CHF patients, and empty circles are the averages over the 54 patients for whom
cardiac death occurred. Top left: $S_E$ in basal condition. Top right: the probability
that basal $S_E$ values from controls and patients were drawn from the same distribution,
evaluated by non parametric test. Bottom left: $S_E$ in paced breathing condition. Bottom
right: the probability that paced breathing $S_E$ values from controls and patients were
drawn from the same distribution, evaluated by non parametric test.\label{fig4}}}
\end{figure}

\begin{figure}[ht!]
\begin{center}
\epsfig{file=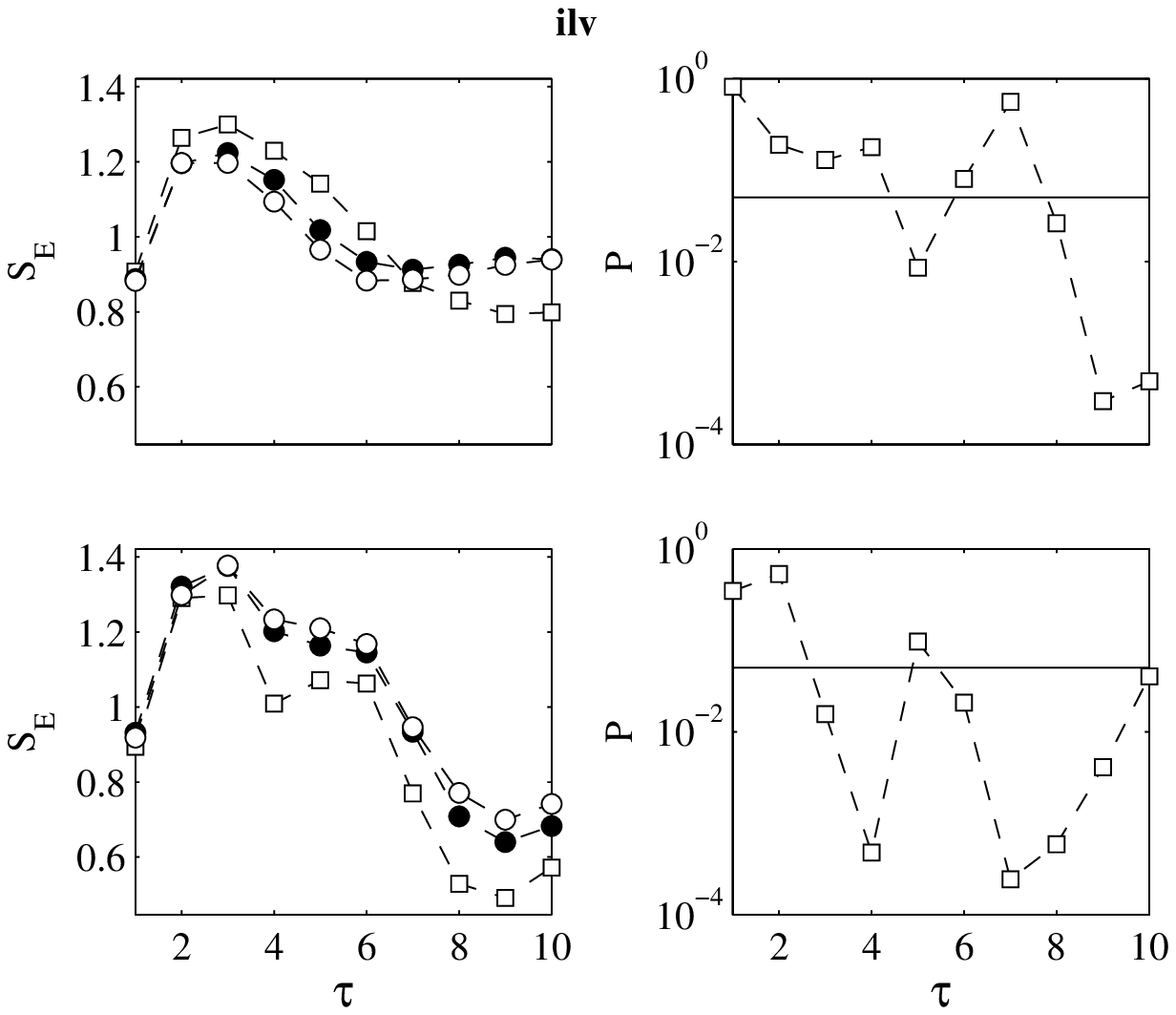,height=8.cm}
\end{center}
\caption{{\small Sample entropy of {\it ilv} time series plotted versus $\tau$. Empty
squares are the averages over the 47 healthy subjects, full circles are the averages over
the 275 CHF patients, and empty circles are the averages over the 54 patients for whom
cardiac death occurred. Top left: $S_E$ in basal condition. Top right: the probability
that basal $S_E$ values from controls and patients were drawn from the same distribution,
evaluated by non parametric test. Bottom left: $S_E$ in paced breathing condition. Bottom
right: the probability that paced breathing $S_E$ values from controls and patients were
drawn from the same distribution, evaluated by non parametric test.\label{fig5}}}
\end{figure}

\begin{figure}[ht!]
\begin{center}
\epsfig{file=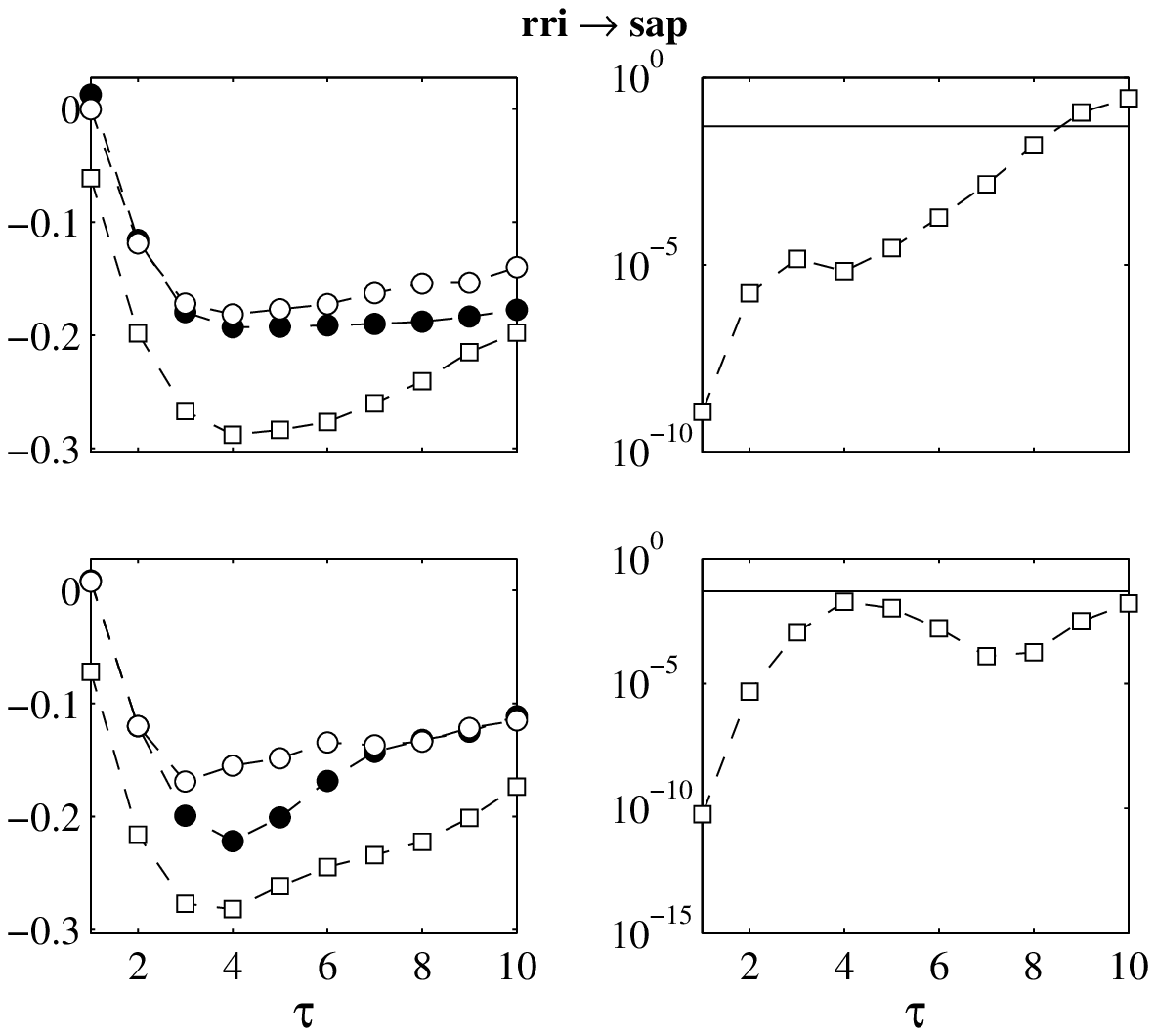,height=8.cm}
\end{center}
\caption{{\small The strength of the interaction {\it rri}$\to${\it sap}, evaluated as
described in the text, is plotted versus $\tau$. Empty squares are the averages over
controls, full circles are the averages over patients, and empty circles are the averages
over dead patients. Top left: {\it rri}$\to${\it sap} in basal condition. Top right: the
probability that basal values from controls and patients were drawn from the same
distribution, evaluated by non parametric test. Bottom left: {\it rri}$\to${\it sap} in
paced breathing condition. Bottom right: the probability that paced breathing values from
controls and patients were drawn from the same distribution, evaluated by non parametric
test.\label{fig6}}}
\end{figure}

\begin{figure}[ht!]
\begin{center}
\epsfig{file=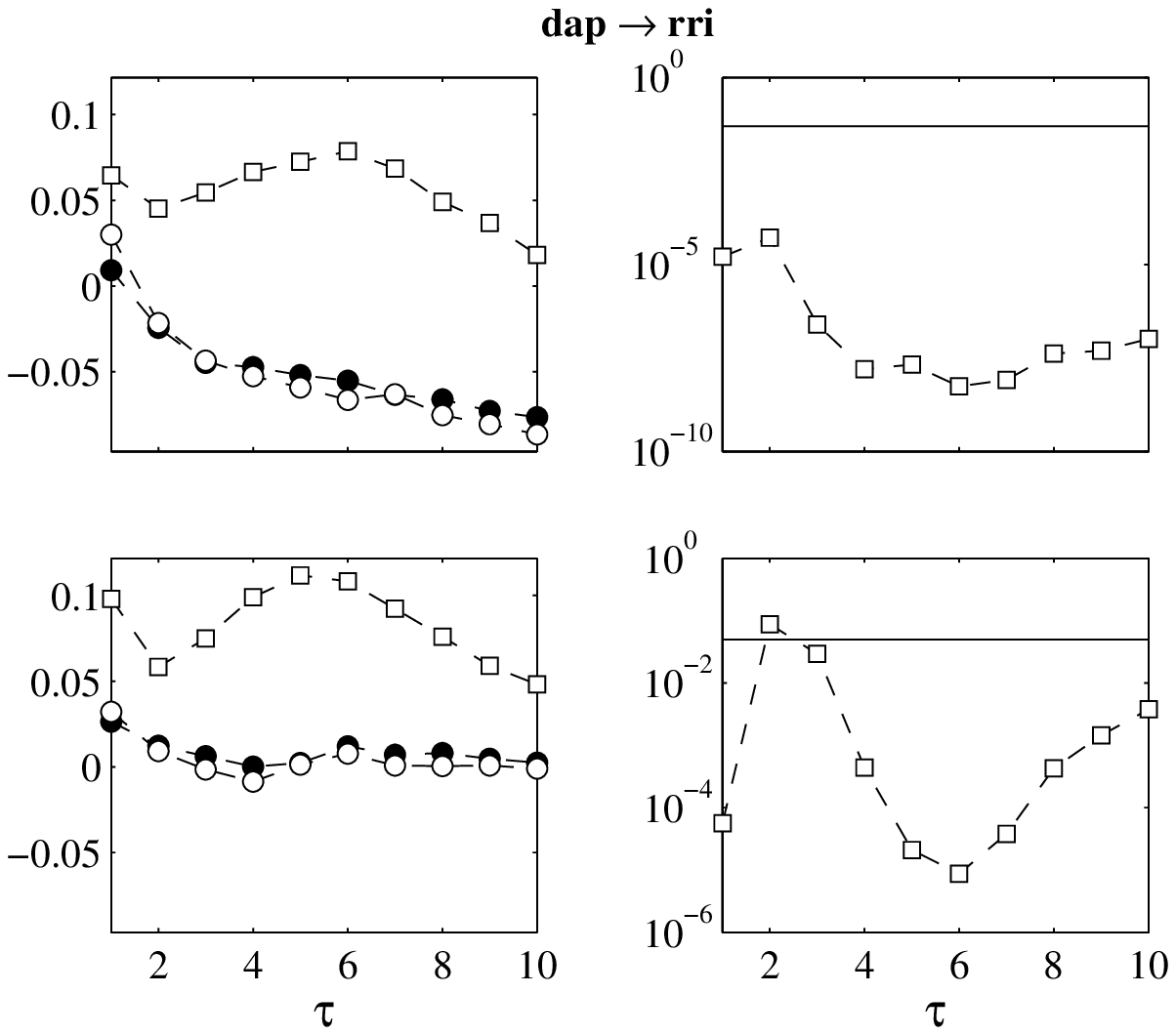,height=8.cm}
\end{center}
\caption{{\small The strength of the interaction {\it dap}$\to${\it rri}, evaluated as
described in the text, is plotted versus $\tau$. Empty squares are the averages over
controls, full circles are the averages over patients, and empty circles are the averages
over dead patients. Top left: {\it dap}$\to${\it rri} in basal condition. Top right: the
probability that basal values from controls and patients were drawn from the same
distribution, evaluated by non parametric test. Bottom left: {\it dap}$\to${\it rri} in
paced breathing condition. Bottom right: the probability that paced breathing values from
controls and patients were drawn from the same distribution, evaluated by non parametric
test.\label{fig7}}}
\end{figure}

\begin{figure}[ht!]
\begin{center}
\epsfig{file=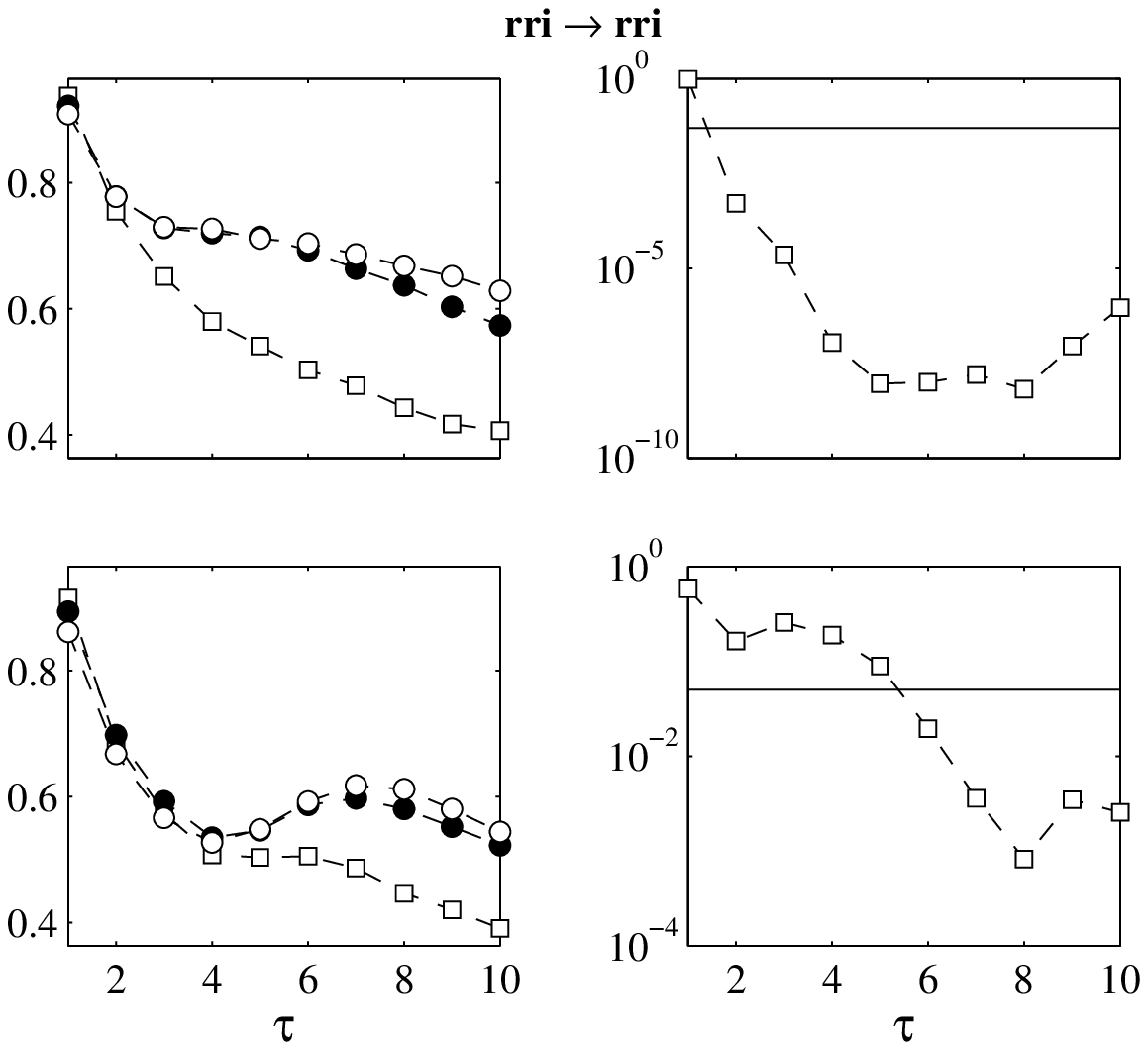,height=8.cm}
\end{center}
\caption{{\small The strength of the interaction {\it rri}$\to${\it rri}, evaluated as
described in the text, is plotted versus $\tau$. Empty squares are the averages over
controls, full circles are the averages over patients, and empty circles are the averages
over dead patients. Top left: {\it rri}$\to${\it rri} in basal condition. Top right: the
probability that basal values from controls and patients were drawn from the same
distribution, evaluated by non parametric test. Bottom left: {\it rri}$\to${\it rri} in
paced breathing condition. Bottom right: the probability that paced breathing values from
controls and patients were drawn from the same distribution, evaluated by non parametric
test.\label{fig8}}}
\end{figure}

\begin{figure}[ht!]
\begin{center}
\epsfig{file=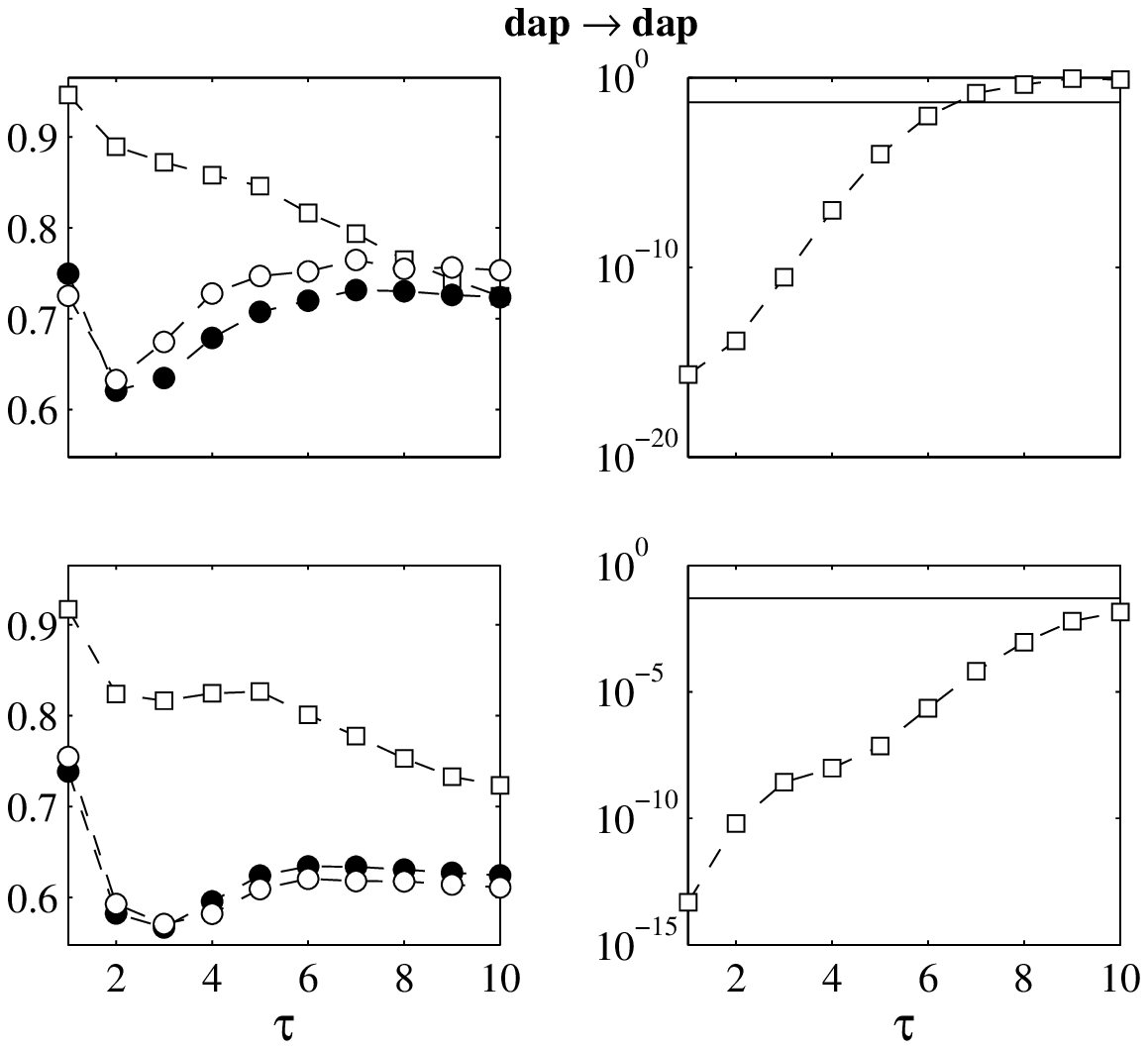,height=8.cm}
\end{center}
\caption{{\small The strength of the interaction {\it dap}$\to${\it dap}, evaluated as
described in the text, is plotted versus $\tau$. Empty squares are the averages over
controls, full circles are the averages over patients, and empty circles are the averages
over dead patients. Top left: {\it dap}$\to${\it dap} in basal condition. Top right: the
probability that basal values from controls and patients were drawn from the same
distribution, evaluated by non parametric test. Bottom left: {\it dap}$\to${\it dap} in
paced breathing condition. Bottom right: the probability that paced breathing values from
controls and patients were drawn from the same distribution, evaluated by non parametric
test.\label{fig9}}}
\end{figure}

\begin{figure}[ht!]
\begin{center}
\epsfig{file=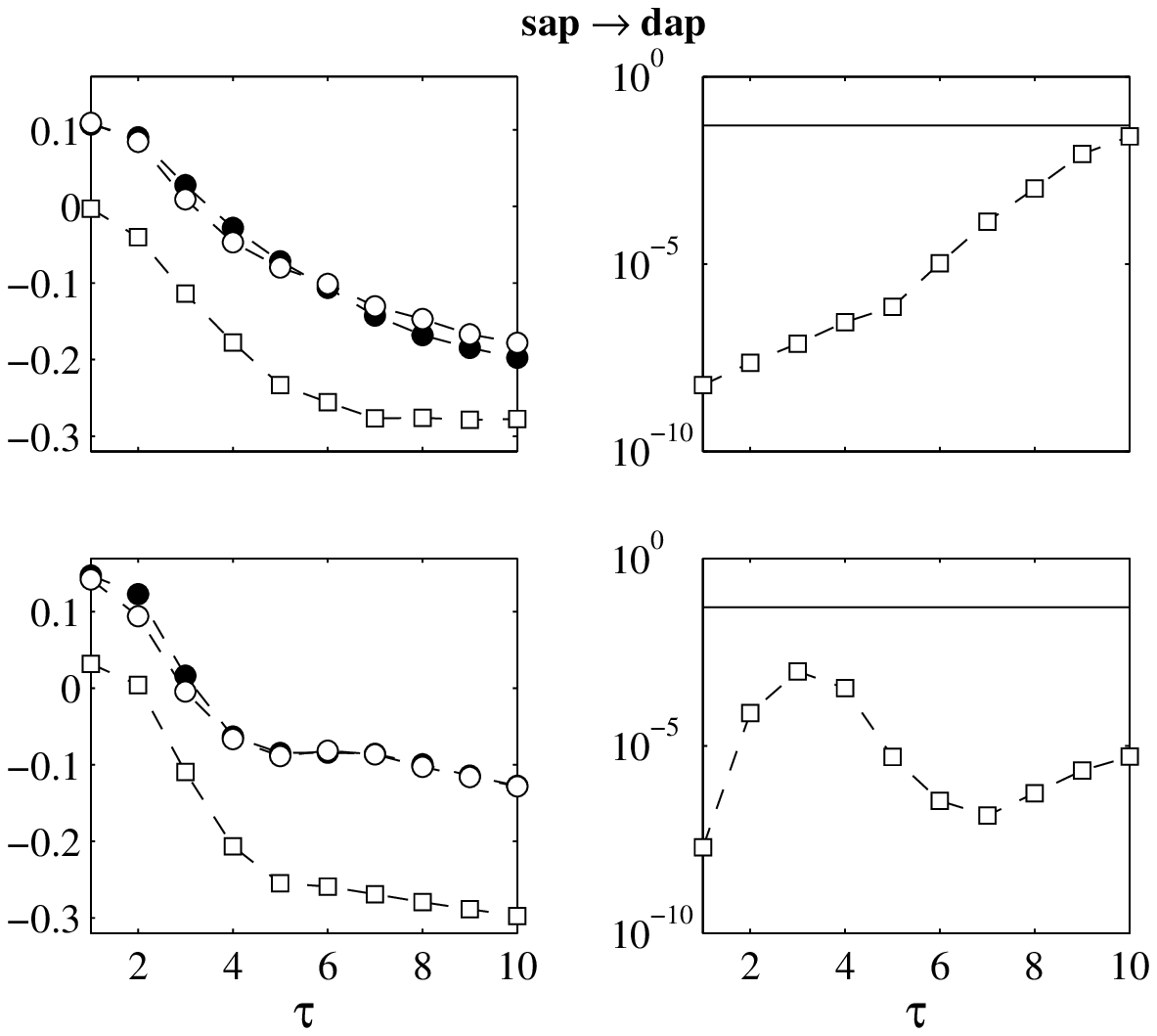,height=8.cm}
\end{center}
\caption{{\small The strength of the interaction {\it sap}$\to${\it dap}, evaluated as
described in the text, is plotted versus $\tau$. Empty squares are the averages over
controls, full circles are the averages over patients, and empty circles are the averages
over dead patients. Top left: {\it sap}$\to${\it dap} in basal condition. Top right: the
probability that basal values from controls and patients were drawn from the same
distribution, evaluated by non parametric test. Bottom left: {\it sap}$\to${\it dap} in
paced breathing condition. Bottom right: the probability that paced breathing values from
controls and patients were drawn from the same distribution, evaluated by non parametric
test.\label{fig10}}}
\end{figure}


\begin{thebibliography}{99}
\bibitem{akselrod}S. Akselrod, D. Gordon, F.A. Ubel, D.C. Shannon and R.J. Cohen, {\it Science} {\bf 213}
220(1981); G.D. Pinna, R. Maestri, G. Raczak, and M.T. La Rovere, {\it Clin Sci (Lond)}
{\bf 103} 81 (2002).
\bibitem{bablo}G.A. Babloyantz, J.M. Salazar and C. Nicolis, {\it Phys. Lett. A} {\bf 111}
152(1985); C.S. Poon, C.K. Merrill, {\it Nature} {\bf 389} 492 (1997).
\bibitem{nunes} L.A. Nunes Amaral, A.L. Goldberger, P.C. Ivanov and H.E. Stanley, {\it Phys. Rev. Lett.} {\bf 81}
2388(1998); Y. Ashkenazy, P.C. Ivanov, S. Havlin, C.K. Peng, A.L. Goldberger and H.E.
Stanley, {\it Phys. Rev. Lett.} {\bf 86} 1900 (2001).
\bibitem{ivanov} P.C. Ivanov, L.A. Nunes Amaral, A.L. Goldberger, S. Havlin, M.G. Rosenblum,
Z. Struzik, and H.E. Stanley, {\it Nature} {\bf 399} 461 (1999); L.A. Nunes Amaral, P.C.
Ivanov, N. Aoyagi, I. Hidaka, S. Tomono, A.L. Goldberger, H.E. Stanley and Y. Yamamoto,
{\it Phys. Rev. Lett.} {\bf 86} 6026 (2001).
\bibitem{lehnertz} K. Lehnertz, C.E. Elger, {\it Phys. Rev. Lett.} {\bf 80} 5019 (1998).
\bibitem{peng}C.K. Peng, J. Mietus, J.M. Hausdorff, S. Havlin, H.E. Stanley and A.L. Goldberger {\it Phys.
Rev. Lett.} {\bf 70} 1343 (1993); P.C. Ivanov, L.A. Nunes Amaral, A.L. Goldberger, S.
Havlin, M.G. Rosenblum, Z. Struzik and H.E. Stanley, {\it Chaos} {\bf 11} 641 (2001).
\bibitem{tass} P. Tass, M.G. Rosenblum, J. Weule, J. Kurths, A. Pikovsky, J. Volkmann, A.
Schnitzler, H-J Freund H-J, {\it Phys. Rev. Lett.} {\bf 81} 3291 (1998).
\bibitem{costa} M. Costa, A.L. Goldberger, C.K. Peng, {\it Phys. Rev. Lett.} {\bf 89} 68102 (2002);
M. Costa, A.L. Goldberger, C.K. Peng, {\it Phys. Rev.} {\bf E 71} 21906 (2005).
\bibitem{complex} J.S. Richman and J.R. Moorman, {\it Am. J. Physiol.} {\bf 278} H2039
(2000); A.L. Goldberger, C.K. Peng, L.A. Lipsitz, {\it Neurobiol. Aging} {\bf 23} 23
(2002).
\bibitem{comment} Vadim V. Nikulin and Tom Brismar
{\it Phys. Rev. Lett.} {\bf 92}, 089803 (2004) ;  M. Costa, A.L. Goldberger, C.K. Peng,
{\it Phys. Rev. Lett.} {\it 92}, 089804 (2004).
\bibitem{paced}S. Rzeczinski, N.B. Janson, A.G. Balanov  and P.V.E. McClintock,
{\it Phys. Rev. E} {\bf 66} 051909 (2002).
\bibitem{ancona}N. Ancona, R. Maestri, D.
Marinazzo, L. Nitti, M. Pellicoro, G.D. Pinna, S. Stramaglia, {\it Physiol. Meas.} {\bf
26} 363 (2005).
\bibitem{kantz} H. Kantz and T. Schreiber, {\it Nonlinear time series
analysis} Cambridge University Press, 1997.
\bibitem{koepchen}{\it Mechanisms of blood pressure waves}, K. Miyakawa, C. Polosa, H.P. Koepchen
(eds.). Springer, Berlin Heidelberg New York (1984).
\bibitem{pin} G.D. Pinna, R. Maestri, S. Capomolla, O. Febo, E. Robbi, F. Cobelli, M.T. La Rovere,
{\it J Am Coll Cardiol} {\bf 46} 1314 (2005).
\bibitem{nollo} G. Nollo, L. Faes, A. Porta, R. Antolini, F. Ravelli, {\it Am. J. Physiol. Heart Circ.
Physiol.} {\bf 288} H1777 (2005).
\bibitem{hirsch}J.A. Hirsch, B. Bishop,{\it Am. J. Physiol.} {\bf 241}, H620 (1981);C. Schafer, M.G. Rosenblum,
H. Abel, {\it Nature} {\bf 392} 239 (1998); C. Schafer, M.G. Rosemblum, H. Abel, J.
Kurths, {\it Phys. Rev.} {\bf E 60} 857 (1999).
\bibitem{winfree} L. Angelini, G. Lattanzi, R. Maestri, D. Marinazzo, G. Nardulli, L. Nitti, M. Pellicoro,
G.D. Pinna, and S. Stramaglia, {\it Phys. Rev.} {\bf E 69}, 061923 (2004).
\bibitem{deboer}R.W. de Boer, J.M.
Karemaker, J. Strackee, {\it Am. J. Physiol.} {\bf 253}, H680 (1987).
\bibitem{swets}J.A. Swets, {\it Science}
{\bf 240} 1285 (1988).
\end{thebibliography}
\end{document}